\documentstyle[a4wide,epsf,11pt,titlepage]{article}

\newcounter{nref}
\newcommand{\bbib}{%
  \renewcommand{\refname}{\large\bf References}%
  \begin{thebibliography}{99}%
  \setcounter{enumiv}{\thenref}}
\newcommand{\ebib}{%
  \end{thebibliography}%
  \setcounter{nref}{\arabic{enumiv}}}
\newcommand{\head}[3]{%
  \setcounter{nref}{0}%
  \thispagestyle{empty}%
  \section*{\LARGE\bf #1}%
  \stepcounter{section}%
  \addcontentsline{toc}{section}{#1}%
  \large\itshape%
  #2\\\vspace{0.1pt}\\%
  #3%
  \normalsize\upshape%
  \bigskip}

\begin{document}


\head{On the Systematics of Core--Collapse Explosions}
     {A.\ Burrows$^1$}
     {$^1$ Department of Astronomy, The University of Arizona, Tucson, AZ 85721}

\subsection*{Abstract}

Recent observations of supernovae, supernova remnants, and radio pulsars
suggest that there are correlations between pulsar kicks and spins,
infrared and gamma-ray line profiles, supernova polarizations, and ejecta debris
fields.  A framework is emerging in which explosion asymmetries play a central
role.  The new perspective meshes recent multi--dimensional theoretical investigations
of the explosion mechanism with trends in $^{56}$Ni yields and explosion 
kinetic energies. These trends imply that the mass above which 
black holes form after collapse is $\sim$30 M$_{\odot}$ and that supernova explosion 
energies may vary by as much as a factor of four.  In addition, new neutrino--matter
opacity calculations reveal that the inner cores of protoneutron stars are
more transparent than hitherto suspected.  This may have consequences for the
delayed neutrino--driven mechanism of explosion itself.  Be that as it may, 
as the millenium dawns a surprising array of new data and
theoretical results are challenging supernova modelers as never before.

\subsection{Introduction}

Summarizing the 
important issues surrounding supernova theory is a daunting task
and fraught with dangers \cite{burrows.1,burrows.2},
but I will attempt here to highlight some of the recent developments
that I find interesting and hope that the reader will be patient
with the manifest limitations of this exercise.  In the process,
some potential systematics will be discussed and new connections
between disparate classes of observations will be suggested.
It is now not unreasonable to imagine a theory that unifies
the spins and velocities of neutron stars, the anisotropies
observed in supernova ejecta, and stellar collapse and explosion.
These may be connected in a given supernova, with the debris asymmetries
correlated with the kick directions and the neutrino and gravitational wave emissions
related to both.  

\subsection{Status of Explosion Modeling}

All groups that do multi--D hydrodynamic modeling of supernovae 
obtain vigorous convection in the semi--transparent mantle bounded by 
the stalled shock \cite{burrows.bhf,burrows.hert,burrows.jm,burrows.mezz,burrows.mwm,burrows.lkw}.  There is a consensus
that the neutrinos drive the explosion \cite{burrows.BW} after a delay whose magnitude has yet to
be determined, but that may be between 100 and 1000 milliseconds.  Whether any 
convective motion or hydrodynamic instability is central to the explosion mechanism
is not clear, with five groups \cite{burrows.bhf,burrows.hert,burrows.jm,burrows.mwm,burrows.lkw} voting yes or maybe  
and one group \cite{burrows.mezz} voting no.  The negative vote is from a group that
is taking pains to handle the transport with a minimum of approximations.
However, this group opted to do the transport in 1--D, save the result, and impose this
history on the 2--D calculation, without feedback.  This prescription is suspect,
but so are the prescriptions of all the other groups, which compromised in different
ways. Hence, a definitive calculation in either two and three dimensions has not
yet been performed.  

It should be noted that it is not trivial to diagnose the differences between
the various calculations of the major groups, nor to reproduce the algorithms they employ.   
In this regard,  one should be cautious of facile comparisons that purport to
explain the results of others.  For instance, to ascribe the explosions that some
groups obtain to the use of the ``gray'' approximation says next to nothing.  There    
is no one ``gray'' approximation, though all share the dubious characteristic
of not being multi--energy--group.  There are many implicit spectra and spectral
forms that can be assumed, various flux limiters, a variety of source terms, a
number of algorithms to merge opaque and transparent regimes, different approximations
for the integrals of the Pauli blocking factors, and different cross sections averages, 
to name only the most obvious.  Furthermore, the differences between a multi--group
flux--limited calculation and a full transport calculation can be larger than the
differences between a well--chosen ``non--gray'' calculation and the latter.  The range of possible
sets of choices under the rubric of ``non--gray'' is vast and each set entails painstaking
evaluation.  Therein lies the major problem: it is harder to assemble a ``non--gray''
code that attempts to cover all the limits than to do the problem correctly.   
It is only in an effort to speed up the calculation that an integral approach is attempted
and doing the full problem is always preferable if sufficient computational resources
are available.  In addition, it is more difficult to have confidence in a patchwork
of approximations than to trust a code that incorporates the full equations, though
poor angular, energy, and spatial zoning can severely compromise even an otherwise
virtuous scheme.

\subsection{Many--Body Correlations}

To focus exclusively on numerical and transport matters is frequently to lose
sight of the important issues.  After the ultimate algorithm is implemented,
the results will depend on the initial progenitor models and the microphysics,
in particular the neutrino cross sections.  In this regard, the recent explorations
into the effects of many--body correlations on neutrino--matter opacities
at high densities are germane \cite{burrows.BS,burrows.BSII,burrows.reddy,burrows.yamada}.  Though the
final numbers have not yet been derived, indications are that we have been
overestimating the neutral--current and the charged--current cross sections
above $10^{14}$ gm cm$^{-3}$ by factors of from two to ten, depending upon density and
the equation of state. The many--body corrections increase with density,
decrease with temperature, and for neutral--current scattering are roughly
independent of incident neutrino energy.  Furthermore, 
the spectrum of energy transfers in neutrino scattering is
considerably broadened by the interactions in the medium. An
identifiable component of this broadening comes from the
absorption and emission of quanta of collective modes akin to the
Gamow--Teller and Giant--Dipole resonances in nuclei (zero-sound;
spin sound), with \v{C}erenkov kinematics.  This implies that all scattering
processes may need to be handled with the full energy redistribution
formalism and that $\nu$--matter scattering at high densities can not be 
considered elastic.  One consequence of this reevaluation is that the late--time 
($\ge 500$ milliseconds) neutrino luminosities may be as much as 50\% larger
for more than a second than heretofore estimated.  These luminosities reflect
more the deep protoneutron star interiors than the early post--bounce luminosities
of the outer mantle and the accretion phase.  Since neutrinos
drive the explosion, this may have a bearing on the specifics of the mechanism,
but it is too soon to tell.

\subsection{Systematics}

Unfortunately, theory is not yet adequate to determine the systematics with progenitor
mass of the explosion energies, residue masses, $^{56}$Ni yields, kicks, or, in fact,
almost any parameter of a real supernova explosion.   Despite this, there are hints,
both observational and theoretical, some of which I would like to touch on here. 
The gravitational binding energy ($B.E.$) exterior to a given interior mass is an increasing
function of progenitor mass, ranging at 1.5 M$_{\odot}$ interior mass from about 
$10^{50}$ ergs for a 10 M$_{\odot}$ progenitor to as much as $3\times10^{51}$ ergs for
a 40 M$_{\odot}$ progenitor \cite{burrows.bhf,burrows.ww}.  This large range must affect the viability of
explosion and its energy.  It is not unreasonable to 
conclude, in a very crude way,  that $B.E.$ sets 
the scale for the supernova explosion energy.  When the ``available'' energy exceeds the
``necessary'' binding energy, both very poorly defined quantities at this stage, explosion
is more ``likely.''  However, how does the supernova, launched in the 
inner protoneutron star, know what binding energy it will be called
upon to overcome when achieving larger radii?  Since 
the post--bounce, pre-explosion accretion rate ($\dot{M}$) is a function
of the star's inner density profile, as is the inner $B.E.$, and since a large $\dot{M}$ seems   
to inhibit explosion,  it may be via $\dot{M}$ that $B.E.$, 
at least that of the inner star, is sensed.  Furthermore, 
a neutrino--driven explosion requires a neutrino--absorbing mass and there is more mass
available in the denser core of a more massive progenitor.  One might think that
binding energy and absorbing mass partially compensate or that a more massive progenitor can  
just wait longer to explode, until its binding energy problems
are buried in the protoneutron star and $\dot{M}$ has subsided.  The net effect
in both cases may be similar explosion energies for different progenitors, though the residue mass
could be systematically higher for the more massive stars.   However, if these effects
do not compensate, the fact that binding energy and absorbing mass are increasing functions
of progenitor mass hints that the supernova explosion energy may also be an increasing function of mass.
Since $B.E.$ varies so much along the progenitor continuum, the range in the explosion energy may 
not be small.  Curiously, the amount of $^{56}$Ni produced explosively also depends upon 
the mass between the residue and the radius at which the shock temperature goes below the explosive
Si--burning temperature, a radius that depends upon explosion energy.  Hence, the amount of
$^{56}$Ni produced may also increase with progenitor mass.  Thermonuclear energy only partially
compensates for the binding energy to be overcome, the former being about $10^{50}$ ergs
for every 0.1 M$_{\odot}$ of $^{56}$Ni produced.  

Not all $^{56}$Ni produced need be ejected.  Fallback is possible and whether there is
significant fallback must depend upon the binding energy profile.  Personally, I think that
there is not much fallback for the lighter progenitors, perhaps for masses below 15
M$_{\odot}$, but that there is significant fallback for the heaviest progenitors.  The transition
between the two classes may be abrupt.  I base this surmise on the miniscule binding energies
and tenuous envelopes of the lightest massive stars and on the theoretical prejudice that
the r--process, or some fraction of it, originates in the protoneutron winds that follow
the explosion for the lightest massive stars \cite{burrows.mbc}.  If there were significant fallback,
these winds and their products would be smothered.  

If there is significant fallback, the supernova may be in jeopardy and much of the $^{56}$Ni 
produced will reimplode.   There may be a narrow range of progenitor mass over which 
the supernova is still viable, while fallback is significant and both the mass of $^{56}$Ni {\it ejected}
and the supernova energy are decreasing.  Above this mass range, a black hole may form.  
Hence, both low--mass and high--mass supernova progenitors may have low $^{56}$Ni
yields.  Recently, two Type IIp supernovae have been detected, 
SN1994W \cite{burrows.soll} and SN1997D \cite{burrows.tur},
which have very low $^{56}$Ni yields ($\le 0.0026\, {\rm M}_{\odot}$ and $\le 0.002\, {\rm M}_{\odot}$, 
respectively), long--duration plateaus, and large inferred ejecta masses ($\ge 25 {\rm M}_{\odot}$).  
The estimated explosion energy for SN1997D is a slight $0.4\times10^{51}$ ergs.  (SN1987A's
explosion energy was $1.5\pm0.5 \times10^{51}$ ergs and its $^{56}$Ni yield was 0.07 M$_{\odot}$.)  
These two supernovae may reside in the fallback gap and imply that the black hole cut--off is
near 30 M$_{\odot}$.  

In sum, supernova $^{56}$Ni yields may vary by a factor of $\sim$100 and may peak at
some intermediate progenitor mass, the supernova explosion
energy may vary by a factor of $\sim 4$ and also may peak at some intermediate progenitor mass, 
and the black hole hole cut--off mass may be near 30 M$_{\odot}$.  However, and importantly, whether
real theoretical calculations will bear out these hinted--at systematics is as yet very unclear. 

\subsection{Young Supernovae and Supernova Remnants}

There are many observational indications that supernova explosions are indeed aspherical.
Fabry--Perot spectroscopy of the young
supernova remnant Cas A, formed around 1680 A.D., reveals that its calcium,
sulfur, and oxygen element distributions are clumped
and have gross back--front asymmetries \cite{burrows.law}. No simple shells are seen.
Many supernova remnants, such as N132D, Cas A, E0102.2-7219, and SN0540-69.3, have systemic
velocities relative to the local ISM of up to 900 km s$^{-1}$ \cite{burrows.kirsh}.
X--ray data taken by ROSAT of the Vela remnant reveal bits of shrapnel with bow shocks \cite{burrows.strom}.
The supernova, SN1987A, is a case study in asphericity: 1) its X--ray, gamma--ray, and optical
fluxes and light curves require that
shards of the radioactive isotope $^{56}$Ni were flung far from the core in
which they were created, 2) the infrared line
profiles of its oxygen, iron, cobalt, nickel, and hydrogen
are ragged and show a pronounced red--blue asymmetry, 3) its light is polarized,
and 4) recent Hubble Space Telescope pictures of its inner debris
reveal large clumps and hint at a preferred direction \cite{burrows.pun}.
Furthermore, radio pictures of the supernova SN1993J, which also has polarized
optical spectral features, depict a broken shell.
One of the most intriguing recent finds is the supernova SN1997X, which is
a so--called Type Ic explosion.  This supernova shows the greatest optical polarization of any to date
(Lifan Wang, private communication).  Type Ic supernovae are thought to be explosions
of the bare carbon/oxygen cores of massive star progenitors stripped of their envelopes.
As such, SN1997X's large polarization implies that the inner supernova cores,
and, hence, the explosions themselves, are fundamentally asymmetrical.
No doubt, instabilities in the outer envelopes of supernova
progenitors clump and mix debris clouds and
shatter spherical shells. The observation of hydrogen
deep in SN1987A's ejecta \cite{burrows.wooden} strongly suggests the work of such
mantle instabilities.  However, the data collectively,
particularly for the heavier elements
produced in the inner core, are pointing to
asymmetries in the central engine of explosion itself.

\subsection{Neutron Star Kicks}

Strong evidence that neutron stars experience a net kick at birth has been mounting
for years.  In 1993 \cite{burrows.harr,burrows.lyne}, it was demonstrated that the pulsars are
the fastest population in the galaxy ($<\!v\!>$ $\sim450$ km s$^{-1}$).
Such speeds are far larger than can result generically from orbital motion
due to birth in a binary (the ``so--called'' Blaauw effect).  An extra ``kick'' is required, probably during
the supernova explosion itself \cite{burrows.fryer}.   In the pulsar binaries, PSR J0045-7319 and PSR 1913+16,
the spin axes and the orbital axes are misaligned, suggesting that the explosions
that created the pulsars were not spherical \cite{burrows.wass,burrows.kaspi}.  In fact, for the former the orbital
motion seems retrograde relative to the spin \cite{burrows.lai} and the explosion may have kicked
the pulsar backwards.  In addition, the orbital eccentricities of Be star/pulsar
binaries are higher than one would expect from a spherical explosion, also implying an
extra kick \cite{burrows.vanden}.  Furthermore, low--mass X--ray binaries (LMXB) are bound neutron star/low-mass
star systems that would have been completely disrupted during the supernova explosion that left
the neutron star, had that explosion been spherical \cite{burrows.kalo}. 
In those few cases, a countervailing kick may have been required to keep the system bound.
The kick had to act on a timescale
shorter than the orbit period and the explosion orbit crossing time.  
Otherwise, the process would have been uselessly adiabatic.  One is 
tempted to evoke as further proof the fact that pulsars
seen around young (age $\le 10^{4}$ years) supernova remnants are on average 
far from the remnant centers,  but here ambiguities 
in the pulsar ages and distances and legitimate questions
concerning the reality of many of the associations make this argument
rather less convincing \cite{burrows.cara,burrows.frail}.  However, 
the ROSAT observations of the 3700 year--old supernova remnant Puppis A 
show an X--ray spot that has been interpreted as its neutron star \cite{burrows.petre}.  This object
has a large X--ray to optical flux ratio, but no
pulsations are seen.  If this interpretation is legitimate, then the inferred neutron star transverse
speed is $\sim$1000 km s$^{-1}$.  Interestingly, the spot is opposite to the position
of the fast, oxygen--rich knots, as one might expect in some models of neutron star
recoil during the supernova explosion.   Whatever the correct interpretation 
of the Puppis A data, it is clear that many neutron stars are given
a hefty extra kick at birth (though the distribution of these kicks 
is broad) and that it is reasonable to implicate asymmetries in the supernova explosion itself.

\subsection{Theories of Kicks}

Supernova theorists have determined that protoneutron star/supernova cores are indeed
grossly unstable to Rayleigh--Taylor--like instabilities \cite{burrows.bhf,burrows.hert,burrows.jm}.  During the 
post--bounce delay to explosion that might last 100 to 1000 milliseconds, 
these cores with 100-- to 200--kilometer radii are strongly convective, boiling and 
churning at sonic ($\sim 3\times10^{4}$ km s$^{-1}$) speeds.  
Any slight asymmetry in collapse can amplify this jostling and result in vigorous
kicks and torques \cite{burrows.bhf,burrows.bh,burrows.spruit} to the residue that can be either systematic or stochastic.   
Whatever the details, it would seem odd if the nascent neutron star 
were not left with a net recoil and spin, though whether pulsar speeds as high
as 1500 km s$^{-1}$ ({\it cf.} the Guitar Nebula) can be reached through this 
mechanism is unknown.  Furthermore, asymmetries in the 
matter field may result in asymmetries in the emission of the neutrinos
that carry away most of the binding energy of the neutron star.
A net angular asymmetry in the neutrino radiation of only 1\%
would give the residue a recoil of $\sim$300 km s$^{-1}$. Not surprisingly, 
many theorists have focussed on producing such a net asymmetry in the
neutrino field, either evoking anisotropic accretion, exotic neutrino flavor physics, or the influence
of strong magnetic fields on neutrino cross sections and transport.  The latter is particularly interesting,
but generally requires magnetic fields of $10^{14}$ to $10^{16}$ gauss \cite{burrows.qian}, far larger than
the canonical pulsar surface field of $10^{12}$ gauss.  Perhaps, the pre--explosion
convective motions themselves can generate via dynamo action the required fields.
Perhaps, these fields are transient and subside to the observed fields after
the agitation of the explosive phase.  It would be hard to hide large fields of $10^{15}$ gauss in the
inner core of an old neutron star, while still maintaining standard surface fields
of $10^{12}$ gauss.   In this context, it is interesting to 
note that surface fields as high as $10^{15}$ gauss
are very indirectly being inferred for the so--called soft gamma repeaters \cite{burrows.duncan}, but these
are a very small fraction of all neutron stars.  If such large fields are necessary
to impart, via anisotropic neutrino emission, the kicks observed, then the coincidence 
that Spruit \& Phinney \cite{burrows.spruit} note between the fields needed to enforce slow pre--collapse
rotation and those observed in pulsars after flux freezing amplification   
is of less significance.  

Whether the kick mechanism is hydrodynamic or due to neutrino momentum,
one might expect that the more massive progenitors would give birth to 
speedier neutron stars.  More massive progenitors generally have more massive
cores. If the kick mechanism relies on the anisotropic ejection of matter \cite{burrows.bh},
then for a given explosion energy and degree of anisotropy we might
expect the core ejecta mass and, hence, the dipole component of the 
ejecta momentum to be larger (``$p \sim \sqrt{2ME}$''), resulting in a larger kick.  
The explosion energy itself may also be larger for the more 
massive progenitors, enhancing the effect.  If the mechanism relies on 
anisotropic neutrino emission, the residues of more massive progenitors
are likely to be more massive and have a greater binding energy ($E_{B} \propto M_{NS}^2$) 
to radiate.  Hence, for a given degree of neutrino anisotropy, the impulse
and kick ($\propto E_{B}/M_{NS}$) would be greater.  In either case, despite the primitive nature
of our current understanding of kick mechanisms, given the above arguements it is not unreasonable  
to speculate that the heaviest massive stars might yield the fastest neutron stars.

\subsection*{Acknowledgements}

I would like to acknowledge productive
conversations with F. Thielemann, K. Nomoto, E. M\"uller, 
W. Hillebrandt, R. Hoffman, S. Reddy, M. Prakash,
B. Schmidt, R. Kirshner, P. Pinto, and C. Wheeler,
as well as support from the NSF under grant No.
AST-96-17494.  I would also like to acknowledge
the facilitating environment of the Santa Barbara
Institute for Theoretical Physics, supported
by the NSF under grant No. PHY94-07194.



\bbib
\bibitem{burrows.1} A. Burrows,  to be published in the proceedings
of the 18'th Texas Symposium on Relativisitc Astrophysics, ed. A. Olinto, J. Frieman, \& D. Schramm
(World Scientific Press, 1998).

\bibitem{burrows.2} A. Burrows, to be published in the proceedings
of the 5'th CTIO/ESO/LCO Workshop ``SN1987A: Ten Years Later,'' eds.
M.M. Phillips \& N.B. Suntzeff, held in La Serena, Chile,
February 24--28, 1997.

\bibitem{burrows.bhf} A. Burrows, J. Hayes, \& B.A. Fryxell, Astrophys. J. {\bf 450} (1995) 830.
\bibitem{burrows.hert} M. Herant, W. Benz, J. Hix, C. Fryer, \& S.A. Colgate, Astrophys. J. {\bf 435} (1994) 339.
\bibitem{burrows.jm} H.-T. Janka \& E. M\"uller, Astron. \& Astrophys. {\bf 290} (1994) 496.
\bibitem{burrows.mwm} D.S. Miller, J.R. Wilson, \& R.W. Mayle, Astrophys. J. {\bf 415} (1993) 278.
\bibitem{burrows.mezz} A. Mezzacappa, {\it et al.}, Astrophys. J. {\bf 495} (1998) 911.
\bibitem{burrows.lkw} I. Lichtenstadt, A. Kholkhov, \& J.C. Wheeler, Astrophys. J., (1998) submitted.
\bibitem{burrows.BW} H. Bethe \& J.R. Wilson, Bethe, Astrophys. J. {\bf 295} (1985) 14.
\bibitem{burrows.BS} A. Burrows \& R. Sawyer, Phys. Rev. C, (1998) in press.
\bibitem{burrows.BSII} A. Burrows \& R. Sawyer, Phys. Rev. Letters, (1998) submitted.
\bibitem{burrows.reddy} S. Reddy, M. Prakash, \& J.M. Lattimer, Phys. Rev. D, (1998) in press.
\bibitem{burrows.yamada} S. Yamada, (1998) this conference.
\bibitem{burrows.ww} T.A. Weaver \& S.E. Woosley, Astrophys. J. Suppl. {\bf 101} (1995) 181.
\bibitem{burrows.mbc} G. Mathews, G. Bazan, \& J. Cowan, Astrophys. J. {\bf 391} (1992) 719.
\bibitem{burrows.soll} J. Sollerman, {\it et al.}, Astrophys. J. {\bf 493} (1998) 933.
\bibitem{burrows.tur} M. Turatto, {\it et al.}, Astrophys. J. Letters, (1998) submitted.
\bibitem{burrows.law} S.S. Lawrence, {\it et al.}, Astron. J. {\bf 109} (1995) 2635.
\bibitem{burrows.kirsh} R.P. Kirshner, J.A. Morse, P.F. Winkler, \& J.P. Blair, Astrophys. J. {\bf 342} (1989) 260.
\bibitem{burrows.strom} R. Strom, H.M. Johnston, F. Verbunt \& B. Aschenbach, Nature {\bf 373} (1994) 590.
\bibitem{burrows.pun} C.S.J. Pun, R.P. Kirshner, P.M. Garnavich, \& P.Challis, B.A.A.S. {\bf 191} (1998) 9901.
\bibitem{burrows.wooden} D.H. Wooden, {\it et al.}, Astrophys. J. Suppl. {\bf 88} (1993) 477.
\bibitem{burrows.harr} P.A. Harrison, A.G. Lyne, \& B. Anderson, Mon. Not. R. Astron. Soc. {\bf 261} (1993) 113.
\bibitem{burrows.lyne} A. Lyne \& D.R. Lorimer, Nature {\bf 369} (1994) 127.
\bibitem{burrows.fryer} C. Fryer, A. Burrows, \& W. Benz, Astrophys.\ J., (1998) in press.
\bibitem{burrows.wass} I. Wasserman, J. Cordes, \& D. Chernoff, (1998) in preparation.
\bibitem{burrows.kaspi} V.M. Kaspi, {\it et al.}, Nature {\bf 381} (1996) 584.
\bibitem{burrows.lai} D. Lai, L. Bildsten, \& V.M. Kaspi, Astrophys. J. {\bf 452} (1995) 819.
\bibitem{burrows.vanden} E.P.J. van den Heuvel \& S. Rappaport, in {\it I.A.U. Colloquium 92}, eds. A. Slettebak \& T.D. Snow
(Cambridge Univ. Press), (1987) pp. 291--308.
\bibitem{burrows.kalo} V. Kalogera, Pub. Astron. Soc. Pac. {\bf 109} (1997) 1394.
\bibitem{burrows.cara} P. Caraveo, Astrophys. J. {\bf 415} (1993) L111.
\bibitem{burrows.frail} D.A. Frail, W.M. Goss, \& J.B.Z. Whiteoak, Astrophys. J. {\bf 437} (1994) 781.
\bibitem{burrows.petre} R. Petre, C.M. Becker, \& P.F. Winkler, Astrophys. J. {\bf 465} (1996) L43.
\bibitem{burrows.bh} A. Burrows \&  J. Hayes, Phys. Rev. Lett. {\bf 76} (1996) 352.
\bibitem{burrows.spruit} H. Spruit \& E.S. Phinney, Nature, (1998) in press.
\bibitem{burrows.qian} D. Lai \& Y.-Z. Qian, Astrophys. J., (1998) in press.
\bibitem{burrows.duncan} M. Duncan \& C. Thompson, B.A.A.S. {\bf 191} (1997) 119.08.

\ebib
\end{document}